\documentclass[aps,prb,onecolumn,floatfix]{revtex4}

\usepackage{epsf}
\usepackage{subfigure}
\usepackage{amsmath}
\usepackage{amssymb}
\usepackage{bm}
\usepackage{graphicx}
\usepackage{epstopdf}
\newcommand{\be}{\begin{equation}}
\newcommand{\ee}{\end{equation}}
\newcommand{\bea}{\begin{eqnarray}}
\newcommand{\eea}{\end{eqnarray}}

\begin{document}

EPL {\bf 103} (2013), 30004

\vskip 1 cm

\title{\bf Information erasure in copolymers}

\author{D. Andrieux\footnote{Presently at Sopra Banking Software} and P. Gaspard}
\affiliation{Center for Nonlinear Phenomena and Complex Systems,
Universit\'e Libre de Bruxelles - Code Postal 231, Campus Plaine, B-1050 Brussels, Belgium}

\begin{abstract}
Information erasure at the molecular scale during the depolymerization of copolymers is shown to require a minimum entropy production in accordance with Landauer's principle and as a consequence of the second law of thermodynamics.  This general result is illustrated with an exactly solvable model of copolymerization, which also shows that the minimum entropy production that is possible for a specific molecular mechanism of depolymerization may be larger than the minimum required by Landauer's principle.
\end{abstract}

\maketitle

\section{Introduction} 

Landauer's principle, according to which erasing information dissipates energy \cite{L61,B62,vN66,B73,B82}, plays a central role in the thermodynamics of information processing.  Recently, these topics have been the focus of theoretical and experimental studies in different systems \cite{G04,AGCGJP07,AGCGJP08,AG08EPL,KLZ07,KPVdB07,TSUMS10,EVdB11,BAPCDL12,MJ12,GK13,BS13}.  In the present letter, our purpose is to consider the erasure of the information contained in copolymers undergoing depolymerization. Copolymers form natural supports of information at the molecular scale.  Information can be processed, transmitted, or erased by attachment or detachment of the units composing copolymers.  These physicochemical mechanisms are ruled by the laws of kinetics and thermodynamics.  In previous work, we established the existence of a fundamental link between thermodynamics and the information contained in the sequence of a growing copolymer \cite{AG08,AG09}.  Here, our aim is to show that Landauer's principle is satisfied during the depolymerization of a copolymer.  This result is illustrated in the case of a class of simple models \cite{AG09}.

Depolymerization as well as polymerization is considered as a Markovian stochastic process for a single copolymer in a solution containing monomers of $M$ different species.  The surrounding solution is supposed to be large enough to constitute a reservoir where the concentration of monomers remains constant during the whole process.  The monomers may randomly attach or detach at one end of the copolymer, its other end being inert.  Whether the copolymer grows or depolymerizes is controlled by the concentrations of monomers.  The speed of these processes is also determined by the values of the rate constants of attachment or detachment for every monomer species $m\in\{1,2,...,M\}$.

\section{Stochastic depolymerization of single copolymers} 

For the present considerations, the copolymer is described as a sequence of monomers $\omega=m_1m_2\cdots m_l$ with $m_j\in\{1,2,...,M\}$ and its length $l=\vert\omega\vert$.  The copolymer is supposed to be long enough for the statistical properties of the sequence $\omega$ to be well defined.  The attachment and detachment of a monomer $m=m_{l+1}$ 
\be
m_1m_2\cdots m_{l} \ + \ m \ \underset{W(\omega m,\omega)}{\overset{W(\omega,\omega m)}{\rightleftharpoons}}
  \  m_1m_2\cdots m_lm
\ee
proceeds at the rates $W(\omega,\omega')$ and $W(\omega',\omega)$ with $\omega'=\omega m$.  The time evolution of the probability $P_t(\omega)$ that the copolymer has the sequence $\omega$ at the current time $t$ is ruled by the master equation \cite{McQ67,S76,NP77,H05}
\be
\frac{d P_t(\omega)}{dt} = \sum_{\omega'} \left[
P_t(\omega') \, W(\omega',\omega)
- P_t(\omega) \, W(\omega,\omega')\right] \, .
\label{master}
\ee

Often, copolymers grow or shrink at constant average speed
\be
v = \frac{d\langle l\rangle_t}{dt} = \frac{d}{dt}\sum_\omega \vert\omega\vert \, P_t(\omega)
\label{speed}
\ee
in regimes of steady growth or depolymerization.  In such regimes, the probability to find a sequence $\omega$ at the current time $t$ can be assumed to factorize as
\be
P_t(\omega) \simeq p_t(l) \, \mu_l(\omega)
\label{factorize}
\ee
into the probability that the copolymer has the length $l=\vert\omega\vert$ at the time $t$ and the stationary probability distribution $\mu_l(\omega)$ of its possible sequences, which is normalized according to $\sum_{\omega:\vert\omega\vert=l}\mu_l(\omega)=1$ \cite{CF63JPS,CF63JCP}.  Equation~(\ref{factorize}) is the leading term of an expansion involving extra terms describing what happens around the growing or shrinking end of the copolymer.  Since the probability in these extra terms is concentrated near the end of the copolymer, they do not contribute to the quantities that are defined per monomer by averaging over the total length of the copolymer.  During copolymer growth, the sequence that is synthesized has a composition that is self-generated by the process so that the stationary distribution $\mu_l(\omega)$ is given as the solution of the master equation (\ref{master}) in terms of the monomer concentrations and the rate constants \cite{AG08,AG09}.  In the case of depolymerization, the initial sequence that is introduced in the solution is characterized by an arbitrary probability distribution $\bar{\mu}_l(\omega)$ because the copolymer has been synthesized under conditions different from the ones in the solution under observation.  The initial copolymer may have a Bernoulli or correlated random sequence.  It may also contain a message crypted in a seemingly random sequence.

\begin{figure}[h]
\centerline{\scalebox{0.6}{\includegraphics{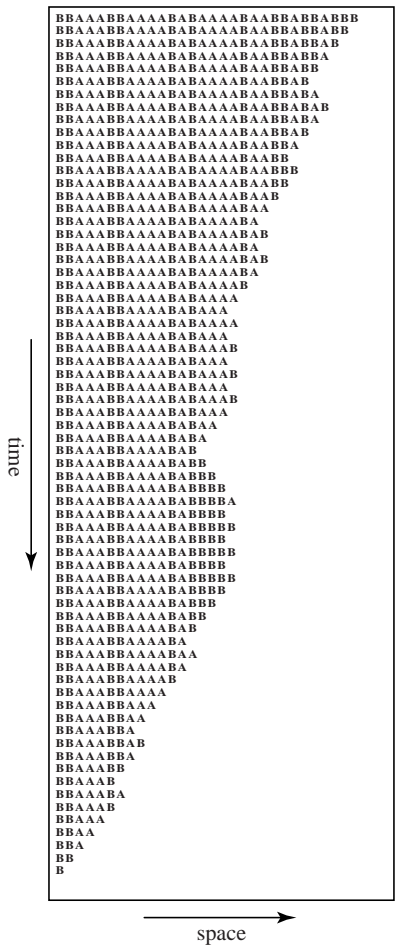}}}
\caption{Space-time plot of the depolymerization process of a random copolymer composed of two monomers A and B generated by a Bernoulli process of equal probabilities $\mu_{\rm A}=\mu_{\rm B}=0.5$.  The transition rates are given by $W(\omega,\omega m)=\kappa_{+m}[m]$ and $W(\omega m,\omega)=\kappa_{-m}$ for $m={\rm A},{\rm B}$ with the rate constants $\kappa_{+{\rm A}}=\kappa_{+{\rm B}}=1$, $\kappa_{-{\rm A}}=10^{-3}$, and $\kappa_{-{\rm B}}=2\times 10^{-3}$.  The concentrations are taken as $[{\rm A}]=3\times 10^{-4}$ and $[{\rm B}]=5\times 10^{-4}$.}
\label{fig1}
\end{figure}

An example of depolymerization is depicted in Fig.~\ref{fig1}.  The initial copolymer is composed of two monomers forming a Bernoulli random sequence with equal probabilities.  The stochastic process is simulated by Gillespie's algorithm \cite{G76,G77}.  The concentrations are such that detachments occur at a higher rate than attachments so that the copolymer shrinks.  We observe that  depolymerization is not monotonous since transient growth phases of different compositions happen at random. Every transient growth phase lasts for a finite time and elongates the copolymer by random short sequences that are eventually decomposed. As a result, and in contrast to  models of information erasure considered in the literature, the sequence to be erased is not fixed, but changes due to the fluctuations in monomer insertions and removals. If the sequence has a finite initial length $l_0$, the depolymerization at the constant average speed (\ref{speed}) stops after a lapse of time $\Delta t\simeq l_0/v$ when the copolymer is completely decomposed.  In order to characterize the dynamical properties in the regime of steady depolymerization, the copolymer is assumed to be arbitrarily long so that decomposition proceeds endlessly.

\section{Thermodynamics of depolymerization} 

For a solution at given pressure and temperature $T$, opposite transition rates are related by
\be
\frac{W(\omega,\omega')}{W(\omega',\omega)} = \exp
\frac{G(\omega)-G(\omega')}{kT} \, ,
\label{ratio}
\ee
where $G(\omega)$ is the free enthalpy or Gibbs free energy of a single copolymer chain $\omega$ and $k$ is Boltzmann's constant \cite{H05}.  The existence of thermodynamic quantities 
associated with a copolymer chain of sequence $\omega$ supposes a separation of time scales between the short time scale of every attachment or detachment event and the long time scale of the copolymer dwelling in the sequence $\omega$ until the next event.   Under this assumption, the enthalpy $H(\omega)$ and the entropy $S(\omega)$ can be defined similarly and they are related by $G(\omega)=H(\omega)-TS(\omega)$.  

Since the copolymer has the probability $P_t(\omega)$ to have the sequence $\omega$ at the current time $t$, its total entropy is given by
\be
S_t = \sum_{\omega} P_t(\omega) S(\omega) - k
\sum_{\omega} P_t(\omega) \ln P_t(\omega) \, , 
\label{entropy}
\ee
where the first term is the statistical average of the entropy $S(\omega)$ of the copolymer with the sequence $\omega$ and the second term is the contribution of the statistical distribution over the different possible sequences $\{\omega\}$ observed at the current time $t$ \cite{G04JCP}.

In this framework, we showed elsewhere \cite{AG08,AG09} that the thermodynamic entropy production is given by the following expression:
\be
\frac{1}{k}\frac{d_{\rm i}S}{dt} = v \left( - \frac{g}{kT} + D\right) \geq 0 \, ,
\label{entr-prod}
\ee
where $v$ is the average speed (\ref{speed}), $g$ is the average free enthalpy per monomer
\be
g = \lim_{l\to\infty} \frac{1}{l} \sum_{\omega:\vert\omega\vert=l} \mu_l(\omega) \,G(\omega) \, ,
\label{g}
\ee
and 
\be
D = \lim_{l\to\infty} -\frac{1}{l} \sum_{\omega:\vert\omega\vert=l} \mu_l(\omega) \,\ln \mu_l(\omega)\geq 0
\label{D}
\ee
is the disorder per monomer in the ensemble of copolymer sequences.
The entropy production (\ref{entr-prod}) is always non-negative by virtue of the second law of thermodynamics.

In the case of depolymerization, the speed is negative, $v<0$, and the non-negativity of entropy production (\ref{entr-prod}) implies that 
\be
g \geq kT\, D \, .
\label{inequal}
\ee
Here, $g$ is the free enthalpy per monomer (\ref{g}) that is consumed to depolymerize the copolymer and $D$ is the disorder per monomer (\ref{D}) in the initial copolymer that is dissolved during depolymerization.  We notice that the inequality (\ref{inequal}) holds between quantities defined in terms of the stationary probability distribution $\mu_l(\omega)$.

{\it Information erasure.} Now, we suppose that the copolymer has initially the given sequence $\bar{\omega}$.  The initial probability is thus given by $P_{t=0}(\omega)=\delta_{\omega,\bar{\omega}}$.  If depolymerization proceeds at a constant speed (\ref{speed}), the probability $\mu_l(\omega)$ takes the unit value for the sequence $\omega=\bar{\omega}$ restricted to the length $l$ and zero otherwise.  The random transient growths of the copolymer contribute by a correction that is concentrated near the end of the copolymer where $P_t(\omega)\neq\delta_{\omega,\bar{\omega}}$.  As long as the extension of these transient growths is finite, they have negligible contributions for the arbitrarily long remaining sequence and the disorder (\ref{D}) is vanishing, $D=0$.  Besides, the free enthalpy per monomer (\ref{g}) takes a value given by the statistical average $\bar{g} = \lim_{l\to\infty} G(\bar{\omega}_l)/l$ where $\bar{\omega}_l$ denotes the restriction of the initial sequence $\bar{\omega}$ to the length $l$.  Therefore, the thermodynamic entropy production is given by
\be
\frac{1}{k}\frac{d_{\rm i}S}{dt} = -v \, \frac{\bar{g}}{kT} \geq 0 \, ,
\label{entr-prod-0}
\ee
where the speed $v$ is negative because the copolymer decomposes. 

On the other hand, an arbitrarily long sequence $\bar{\omega}$ can be characterized by the occurrence frequencies of singlets $m_j$, doublets $m_jm_{j+1}$, triplets $m_jm_{j+1}m_{j+2}$, etc.:
\be
\bar{\mu}_1(m) \, , \ \bar{\mu}_2(mm') \, , \ \bar{\mu}_3(mm'm'') \, , \ldots \, .
\label{frequencies}
\ee
These frequencies define a probability distribution that characterizes all the statistical properties of the copolymer undergoing depolymerization.  In particular, the free enthalpy per monomer $\bar{g}$ can be obtained using Eq.~(\ref{g}) in terms of the frequencies (\ref{frequencies}).  Similarly, the information content of the sequence $\bar{\omega}$ -- measured in nats per monomer -- can be characterized by the Shannon information per monomer given by 
\be
\bar{I}_{\infty} = \lim_{l\to\infty} \bar{I}_{l} \, ,
\label{Info-infty}
\ee
with
\be
\bar{I}_{l} = -\frac{1}{l} \sum_{\omega:\vert\omega\vert=l} \bar{\mu}_l(\omega) \,\ln \bar{\mu}_l(\omega)\geq 0 \, .
\label{Info-l}
\ee
As long as the statistical properties of the single copolymer under depolymerization are fully characterized by the probability distribution $\bar{\mu}_l(\omega)$, its free enthalpy per monomer has the same value as in the corresponding statistical ensemble where the copolymers have a disorder per monomer equal to the quantity (\ref{Info-infty}). But, we have just shown with Eq.~(\ref{inequal}) that the free enthalpy per monomer is bounded from below by the disorder per monomer multiplied by the thermal energy $kT$:
\be
\bar{g} \geq kT\, \bar{I}_{\infty} \, .
\label{inequal-0}
\ee
Therefore, the depolymerization of a copolymer $\bar{\omega}$ containing an amount of information equal to $\bar{I}_{\infty}$ nats per monomer requires the minimum dissipation rate
\be
\frac{1}{k}\frac{d_{\rm i}S}{dt} \geq -v \, \bar{I}_{\infty} \geq 0 \, ,
\label{min_dissip}
\ee
which is the manifestation of Landauer's principle at molecular scale \cite{L61,B62,vN66,B73,B82,AG08EPL}.  We notice that Eq.~(\ref{min_dissip}) constitutes a general lower bound on the dissipation required to decompose the copolymer. Accordingly, a thermodynamic efficiency of depolymerization can be introduced as
\be
0 \leq \eta \equiv \frac{kT\, \bar{I}_{\infty}}{\bar{g}} \leq 1 \, .
\label{eff}
\ee
For a given depolymerization mechanism, the minimum dissipation is actually determined by the dependence of the reaction rates on the sequence at the end of the copolymer, as illustrated in the following example.

\begin{figure*}[t]
\centerline{\scalebox{0.4}{\includegraphics{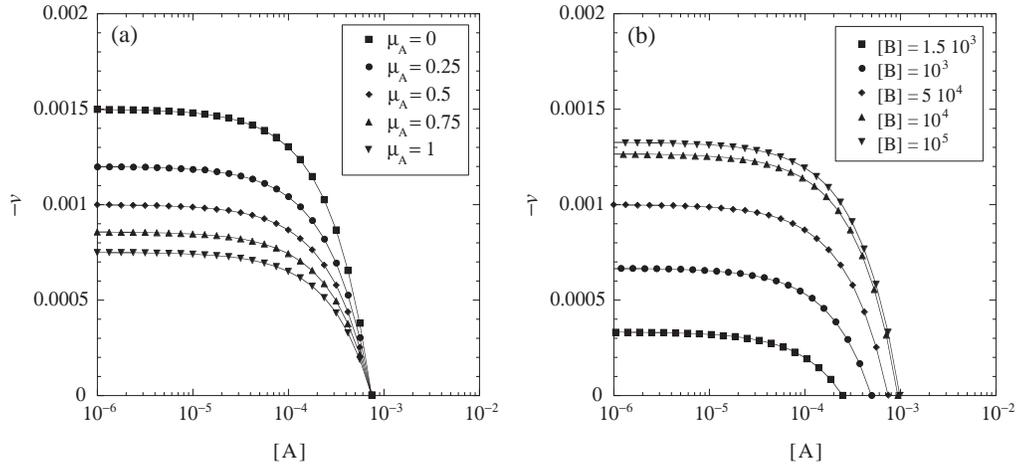}}}
\caption{Depolymerization speed versus the concentration $[{\rm A}]$: (a)~for Bernoulli sequences of different probabilities $\mu_{\rm A}$ at the concentration $[{\rm B}]=5\times 10^{-4}$; (b)~for different values of the concentration $[{\rm B}]$ at the probability $\mu_{\rm A}=0.5$. The initial copolymer is generated as a Bernoulli sequence of probabilities $(\mu_{\rm A},\mu_{\rm B}=1-\mu_{\rm A})$ and length $l=5\times10^5$. The rate constants have the values $\kappa_{+{\rm A}}=\kappa_{+{\rm B}}=1$, $\kappa_{-{\rm A}}=10^{-3}$, and $\kappa_{-{\rm B}}=2\times 10^{-3}$.  The symbols depict the results of numerical simulations with Gillespie's algorithm \cite{G76,G77} and the solid lines the predictions of formula (\ref{speed-model}) with the maximal speed (\ref{v_max}).}
\label{fig2}
\end{figure*}

\begin{figure*}[t]
\centerline{\scalebox{0.49}{\includegraphics{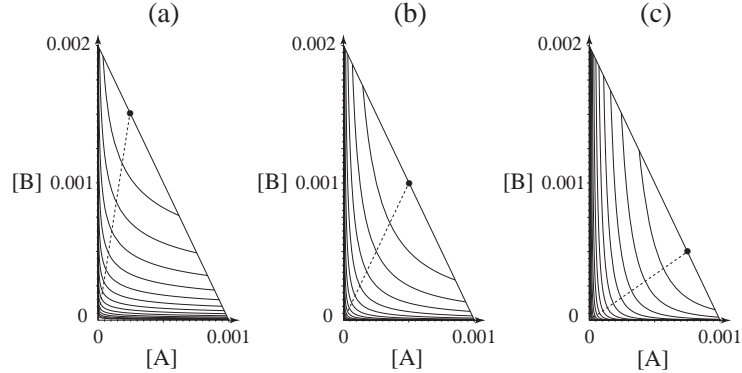}}}
\caption{Contour plot in the concentration space $([{\rm A}],[{\rm B}])$ of the free enthalpy per monomer dissipated during depolymerization with the rate constants $\kappa_{+{\rm A}}=\kappa_{+{\rm B}}=1$, $\kappa_{-{\rm A}}=10^{-3}$, and $\kappa_{-{\rm B}}=2\times 10^{-3}$ for a copolymer of composition (a)~$\bar{\mu}_{\rm A}=0.25$; (b)~$\bar{\mu}_{\rm A}=0.5$; (c)~$\bar{\mu}_{\rm A}=0.75$. The dot is the point (\ref{conc_opt-0}) of minimum dissipation $\bar{g}=kT \bar{I}_1$.  The contour lines are taken at the values $\bar{g}=x kT \bar{I}_1$ with $x=1.5,2,2.5,3,...$.  The dashed lines show the points (\ref{conc_opt}) where the dissipation is minimum at fixed value of the speed $v$.}
\label{fig3}
\end{figure*}

\section{Simple model of depolymerization}  

The main features of our result (\ref{min_dissip}) can already be illustrated with a simple model of copolymerization processes \cite{AG09}.  This model supposes that the attachment of monomer species $m\in\{1,2,...,M\}$ proceeds at the transition rate $W(\omega,\omega m)=\kappa_{+m}[m]$, where $[m]$ denotes the concentration of the monomer $m$ in the surrounding solution, and the corresponding detachment at the rate $W(\omega m,\omega)=\kappa_{-m}$.  
By Eq.~(\ref{ratio}), the free enthalpies of the sequences $\omega$ and $\omega m$ are related by
\be
G(\omega m) = G(\omega) - kT\, \ln \frac{\kappa_{+m}[m]}{\kappa_{-m}} \, .
\ee
Since the rates of the present model only involve the last monomer that is detached or attached to the copolymer, the average free enthalpy per monomer is given by
\be
\bar{g} = - kT \sum_m \bar{\mu}_1(m) \ln \frac{\kappa_{+m}[m]}{\kappa_{-m}} \, ,
\label{g-model}
\ee
which is non-negative $\bar{g}\geq 0$ if the surrounding solution decomposes the chain and erases its information content.  The decomposition speed is given by
\be
v = - v_{\rm max} \left( 1 - \sum_m \frac{\kappa_{+m}[m]}{\kappa_{-m}}\right) \leq 0 \, ,
\label{speed-model}
\ee
where
\be
v_{\rm max} = \frac{1}{\sum_m \left[\bar{\mu}_1(m)/\kappa_{-m}\right]}
\label{v_max}
\ee
is the absolute value of the maximal decomposition speed, achieved when all the concentrations vanish, $[m]=0$.  

The formula (\ref{speed-model}) is derived as follows.  Fluctuations make the copolymer randomly shrink or grow.  Starting from a given monomer $m$, the growth phase will last for a random duration $T_m$ before the copolymer shrinks enough to remove that specific monomer.  The important observation is that future growth or shrinking phases are independent once this original monomer is removed.  Therefore, the speed will be given by
\be
v = - \frac{1}{\sum_m \bar{\mu}_1(m) \langle T_m\rangle}
\label{speed-mean_time} \, ,
\ee
where $\langle T_m\rangle$ is the mean time before a given monomer $m$ at a given location is removed from the chain.   This mean time is obtained by a first-passage calculation \cite{KT75,KT81} giving
\be
\langle T_m\rangle =  \frac{1}{\kappa_{-m}\left( 1 - \sum_{m'} \frac{\kappa_{+m'}[m']}{\kappa_{-m'}}\right)} \, .
\label{mean_time}
\ee
Inserting this expression into Eq.~(\ref{speed-mean_time}) leads to the formula~(\ref{speed-model}) with Eq.~(\ref{v_max}).  To decompose the chain and erase its information content, the concentrations of the monomers should thus satisfy the condition
\be
\sum_m \frac{\kappa_{+m}[m]}{\kappa_{-m}} <1
\label{depolym}
\ee
so that the speed (\ref{speed-model}) is negative. The free enthalpy per monomer (\ref{g-model}) is then indeed non-negative because $\ln(\kappa_{+m}[m]/\kappa_{-m})<0$ for every $m\in\{1,2,...,M\}$.

Figure~\ref{fig2} shows the agreement between the predictions of Eq.~(\ref{speed-model}) and numerical simulations for depolymerization of a copolymer composed of two monomers, $m=1\leftrightarrow{\rm A}$ and $m=2\leftrightarrow{\rm B}$.  We also observe that the decomposition speed vanishes at the limit of the condition (\ref{depolym}).  This limit does not depend on the composition, as seen in Fig.~\ref{fig2}a.  The maximal speed is asymptotically reached for vanishing concentrations, as Fig.~\ref{fig2}b demonstrates.

The free enthalpy per monomer (\ref{g-model}) that is consumed during depolymerization is depicted in Fig.~\ref{fig3} as a contour plot in the plane of the concentrations $([{\rm A}],[{\rm B}])$ for three different initial compositions $\bar{\mu}_m=\bar{\mu}_1(m)$ of the copolymer.  Depolymerization occurs if $\kappa_{+{\rm A}}[{\rm A}]/\kappa_{-{\rm A}}+\kappa_{+{\rm B}}[{\rm B}]/\kappa_{-{\rm B}} < 1$.  The minimum dissipation is  given by $\bar{g} = kT \bar{I}_1$,
where 
\be
\bar{I}_1 = - \sum_m \bar{\mu}_1(m) \ln \bar{\mu}_1(m) \geq 0 
\label{Info-1}
\ee
is the information per monomer (\ref{Info-l}) determined by the frequencies $\bar{\mu}_1(m)$ of singlets in the sequence.  We notice that $\bar{I}_1\geq\bar{I}_{\infty}$ because the information (\ref{Info-infty}) takes a lower value than $\bar{I}_1$ if there are statistical correlations between successive monomers in the sequence. 
The minimum dissipation $\bar{g} = kT \bar{I}_1$ is reached for the concentrations
\be
[m] = \frac{\kappa_{-m}}{\kappa_{+m}}\, \bar{\mu}_1(m)
\label{conc_opt-0}
\ee 
marked by a dot in Fig.~\ref{fig3}. At these concentrations, the decomposition speed vanishes as well as the thermodynamic entropy production (\ref{entr-prod-0}).

\begin{figure}[h]
\centerline{\scalebox{0.45}{\includegraphics{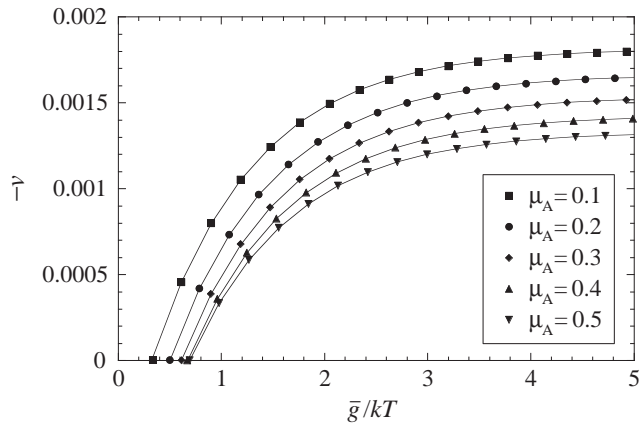}}}
\caption{Absolute value of the fastest decomposition speed versus the dissipation per monomer $\bar{g}/kT$ during depolymerization with the rate constants $\kappa_{+{\rm A}}=\kappa_{+{\rm B}}=1$, $\kappa_{-{\rm A}}=10^{-3}$, and $\kappa_{-{\rm B}}=2\times 10^{-3}$. The initial copolymers are generated as Bernoulli sequences of probabilities $(\mu_{\rm A},\mu_{\rm B}=1-\mu_{\rm A})$ and initial lengths $l=5\times 10^5$. Plotted data are taken at the concentrations (\ref{conc_opt}) linked by  $\frac{\kappa_{+{\rm A}}[{\rm A}]}{\kappa_{-{\rm A}}\mu_{\rm A}}=\frac{\kappa_{+{\rm B}}[{\rm B}]}{\kappa_{-{\rm B}}\mu_{\rm B}}\geq 1$.  Minimum dissipation occurs at $\bar{I}_1\simeq 0.325, 0.500, 0.611, 0.673, 0.693$ for $\mu_{\rm A}=0.1, 0.2, 0.3, 0.4, 0.5$, respectively.  The symbols are the results of numerical simulations with Gillespie's algorithm \cite{G76,G77} and the solid lines the prediction of Eq.~(\ref{v_opt}).}
\label{fig4}
\end{figure}

The minimum of the free enthalpy per monomer (\ref{g-model}) for a fixed decomposition speed $v$ is obtained by using a Lagrange multiplier, which gives
\be
\bar{g}^* =kT \, \bar{I}_1 - kT \, \ln\left( 1 + \frac{v}{v_{\rm max}}\right) \, .
\label{g-star}
\ee 
This value is reached at the concentrations
\be
[m] = \frac{\kappa_{-m}}{\kappa_{+m}}\, \bar{\mu}_1(m) \left( 1 + \frac{v}{v_{\rm max}}\right)
\label{conc_opt}
\ee 
marked by a dashed line in Fig.~\ref{fig3}.

\begin{figure}[h]
\centerline{\scalebox{0.45}{\includegraphics{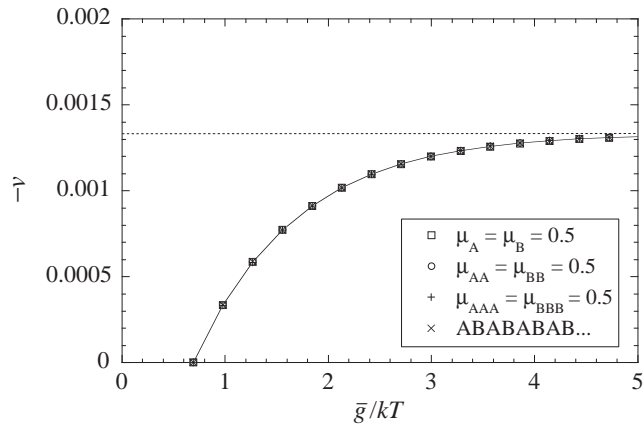}}}
\caption{Absolute value of the decomposition speed versus the dissipation per monomer $\bar{g}/kT$ during depolymerization with the rate constants $\kappa_{+{\rm A}}=\kappa_{+{\rm B}}=1$, $\kappa_{-{\rm A}}=10^{-3}$, and $\kappa_{-{\rm B}}=2\times 10^{-3}$ for different initial copolymers all with the same frequencies $\bar{\mu}_{\rm A}=\bar{\mu}_{\rm B}=0.5$ and singlet information $\bar{I}_1=\ln 2$, but generated as a Bernoulli sequence composed of singlet probabilities $\mu_{\rm A}=\mu_{\rm B}=0.5$ and information $\bar{I}_{\infty}=\ln 2$ (open squares),  of doublet probabilities $\mu_{\rm AA}=\mu_{\rm BB}=0.5$ and information $\bar{I}_{\infty}=(\ln 2)/2$ (open circles), of triplet probabilities $\mu_{\rm AAA}=\mu_{\rm BBB}=0.5$ and information $\bar{I}_{\infty}=(\ln 2)/3$ (pluses), and as a periodic sequence ABABABAB... of information $\bar{I}_{\infty}=0$ (crosses). The minimum dissipation is determined by $\bar{I}_1=\ln 2$ for every copolymer. The symbols are the results of numerical simulations with Gillespie's algorithm \cite{G76,G77}, the solid line is the prediction of Eq.~(\ref{v_opt}), and the dashed line the asymptotic maximum speed (\ref{v_max}).}
\label{fig5}
\end{figure}

Reciprocally, for a fixed free-enthalpy consumption $\bar{g}$, the fastest erasure speed is given by
\be
v^* = -v_{\rm max} \left[1- \exp\left(\bar{I}_1-\frac{\bar{g}}{kT}\right)\right] \, ,
\label{v_opt}
\ee
which is obtained by inverting Eq.~(\ref{g-star}). This speed vanishes at the minimum dissipation $\bar{g}=kT \bar{I}_1$ and approaches the maximal speed $-v_{\rm max}$ exponentially as the allowed dissipation increases.  This behavior is seen in Fig.~\ref{fig4} where a comparison is made between the result (\ref{v_opt}) and numerical simulations at the concentrations (\ref{conc_opt}).  We observe that the threshold of minimum dissipation is indeed given by the information (\ref{Info-1}).  Figure~\ref{fig5} plots the speed versus the dissipation for different sequences of the two monomers A and B, but all with the same frequencies $\bar{\mu}_{\rm A}$ and $\bar{\mu}_{\rm B}$.  The singlet information per monomer (\ref{Info-1}) is the same for the different sequences, although the information (\ref{Info-infty}) differs among them.  Since $\bar{I}_1\geq\bar{I}_{\infty}$, Eq.~(\ref{min_dissip}) constitutes a general lower bound, which may not be reached for specific depolymerization mechanisms. For a molecular mechanism involving only singlets, as in the present model, the efficiency (\ref{eff}) is maximal in the limit of vanishing decomposition speed where it takes the value
\be
\eta_0 = \frac{\bar{I}_{\infty}}{\bar{I}_1} \leq 1 \, ,
\ee
for a given copolymer characterized by the information (\ref{Info-infty}) contained in the whole sequence and the information (\ref{Info-1}) in the frequencies $\bar{\mu}_1(m)$ of the singlets. At finite erasure speed $v$, the maximal efficiency is given by $\eta^*(v)=kT\bar{I}_{\infty}/\bar{g}^*(v)$. It decreases with the speed until it vanishes as a cusp located at minus the maximal speed (\ref{v_max}).

\section{Conclusions}

In the present Letter, Landauer's principle is shown to hold at the molecular scale during the depolymerization of copolymers with information contained in their monomer sequence.  
Depolymerization is a physico-chemical process responsible for the erasure of information at the nanoscale of individual molecules.  This process is the reverse of copolymerization, which can generate or transmit information under nonequilibrium conditions \cite{AG08,AG09}. Our analysis shows that the depolymerization of an initial copolymer  with some information content $\bar{I}_{\infty}$ requires in general the minimum of thermodynamic entropy production given by Eq.~(\ref{min_dissip}).  This general result is illustrated with an exactly solvable model of copolymerization.  
For this model, the decomposition speed has been obtained analytically, as well as the free enthalpy dissipated per monomer during depolymerization.  Interestingly, the analysis of this model shows that the minimum dissipation depends on the specific mechanism taking place at the molecular level and may be larger than the general lower bound (\ref{min_dissip}).  More realistic models of copolymerization can be studied by similar methods that we hope to report on in a future publication.

\vskip 0.75 cm
{\bf Acknowledgments.} This research is financially supported by the Universit\'e Libre de Bruxelles and the Belgian Federal Government under the Interuniversity Attraction Pole project P7/18 ``DYGEST".


\end{document}